\documentclass[pre,twocolumn,showpacs,amssymb,floatfix]{revtex4}
\usepackage{graphicx,epsfig}   \usepackage{subfigure} \usepackage{amsmath,
amsfonts, amssymb}

\newcommand{\fig}{Fig.} 
\begin{document}

\title{Passage Times for Unbiased Polymer Translocation through a
  Narrow Pore}

\author{Joanne Klein Wolterink, Gerard T. Barkema and Debabrata Panja}
\affiliation{Institute for Theoretical Physics, Universiteit Utrecht,
Minnaertgebouw, Leuvenlaan 4, Postbus 80.195, 3508 TD Utrecht, The
Netherlands}

\begin{abstract} 
We study the translocation process of a polymer in the absence of
external fields for various pore diameters $b$ and membrane thickness
$L$. The polymer performs Rouse and reptation dynamics. The mean
translocation time $\langle\tau_t\rangle$ that the polymer needs to
escape from a cell, and the mean dwell time $\langle\tau_d\rangle$
that the polymer spends in the pore during the translocation process,
obey scaling relations in terms of the polymer length $N$, $L$ and
$b/R_g$, where $R_g$ is the radius of gyration for the polymer. We
explain these relations using simple arguments based on polymer
dynamics and the equilibrium properties of polymers.
\end{abstract}


\maketitle

Transport of molecules across membranes is an essential mechanism for
life processes. These molecules are often long, and the narrow pores in the
membranes do not allow them to pass through 
as a single unit. They have to thus squeeze
--- i.e., translocate --- themselves through the pores. DNA, RNA and
proteins are such naturally occuring long molecules
\cite{drei,henry,akimaru,goerlich,schatz} in a variety of biological
processes. Translocation is used in gene therapy \cite{szabo,hanss}, 
and in delivery of drug molecules to their activation sites
\cite{tseng,tsutsui}. Understandably, the process of translocation has
been an active topic of current research: not only because it is a
cornerstone of many biological processes, but also due to its
relevance for practical applications.

Translocation is a complicated process in living organisms --- the
presence of chaperon molecules, pH, chemical potential gradients, and
assisting molecular motors strongly influence its
dynamics. Consequently, the translocation process has been empirically
studied in great variety in biological literature
\cite{wickner,simon}. Study of translocation as a \emph{biophysical}
process is more recent. Herein, the polymer is simplified to a
sequentially connected string of $N$ monomers as it passes through a
narrow pore on a membrane. The quantities of interest are the typical
time scale for the polymer to leave a confining cell (the ``escape of
a polymer from a vesicle'' time scale) \cite{sungpark1}, and the typical
time scale the polymer spends in the pore (the ``dwell'' time scale)
\cite{sungpark2} as a function of $N$ and other parameters like
membrane thickness, membrane adsorption, electrochemical potential
gradient, etc.\ \cite{lub}.

These quantities have been measured directly in numerous experiments
\cite{expts}. A number of (mean-field type) theories have been
proposed for the scaling of these typical times
\cite{sungpark1,sungpark2,lub} during the last decade as well. They
describe translocation as a first-passage or Kramer's problem over an
entropic barrier in terms of the ``reaction coordinate'' $m$
alone. Here $m$ is the number of the monomer threaded into the pore
($m=1, \ldots, N$), and the transition rates from $m$ to $m\pm1$ and
vice versa are obtained from the derivatives of the free energy
w.r.t.\ $m$. 

\begin{figure*}
\begin{center}
\begin{minipage}{0.29\linewidth}
\includegraphics[width=0.98\linewidth]{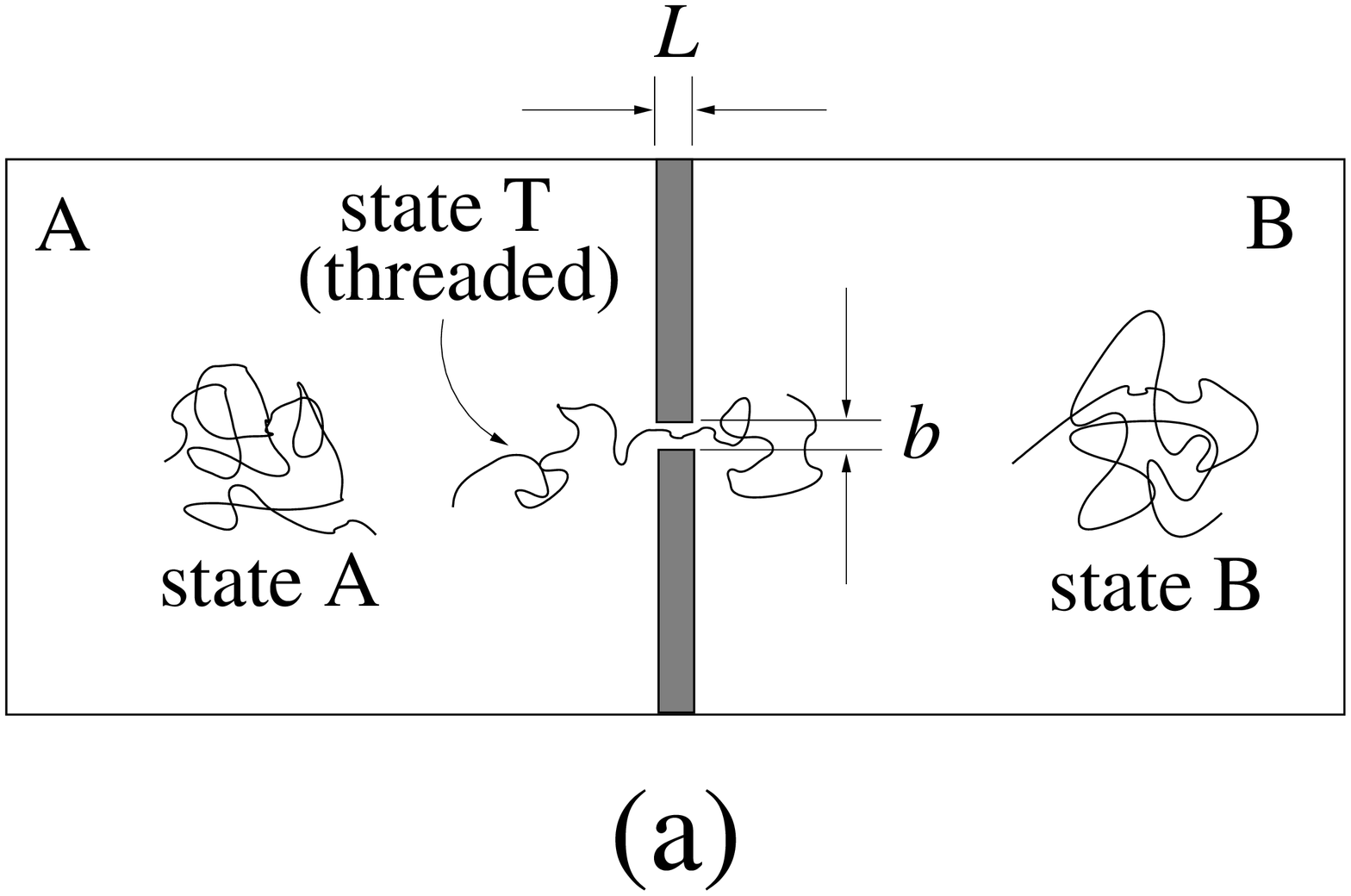}
\end{minipage}
\hspace{5mm}
\begin{minipage}{0.65\linewidth}
\includegraphics[width=0.98\linewidth]{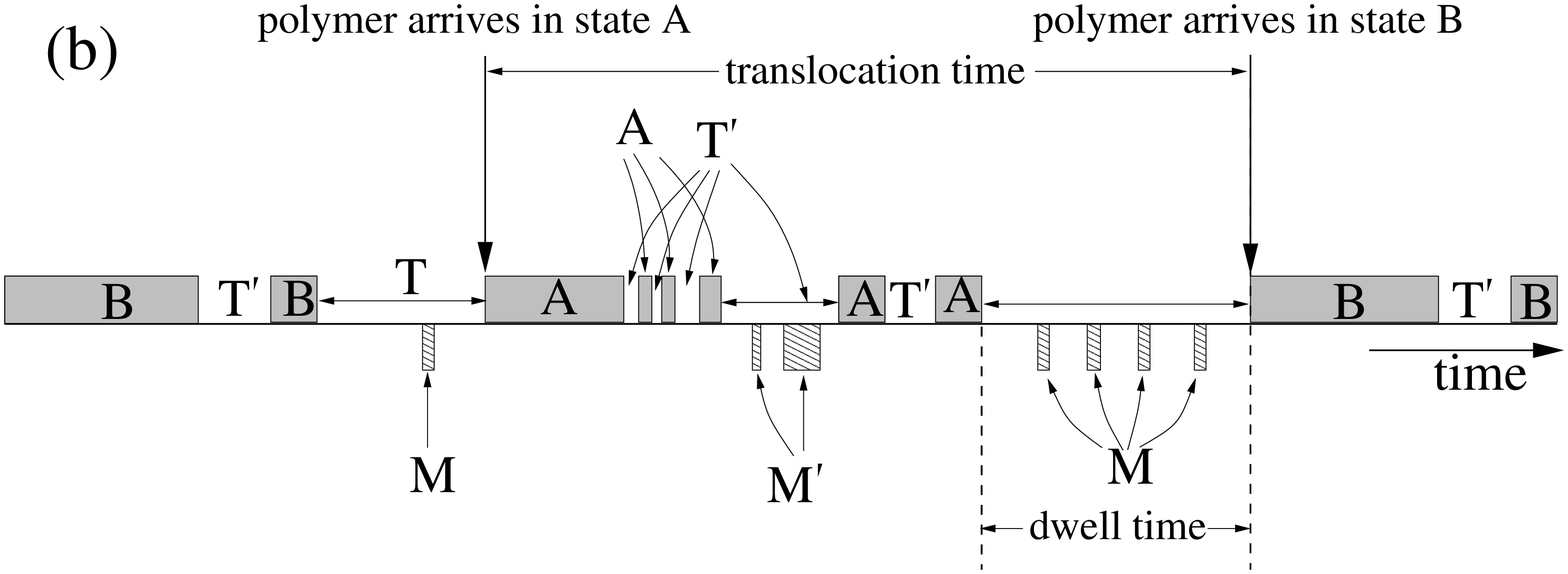}
\end{minipage}
\caption{(a) Our system with polymers in different states. (i) state A:
all monomers are in cell A; (ii) state T (threaded): some monomers are
in cell A and the rest in cell B; (iii) state B: all monomers are in
cell B. (b) A typical translocation process of the polymer [using the
definition of polymer states defined in Fig. \ref{fig2}(a)] for our
system.}
\label{fig2}
\end{center}
\end{figure*}
In the context of unbiased polymer translocation (i.e., in the absence
of external driving fields), the prediction of the mean-field theories
(which only consider polymers with simple random walk statistics) is
that the dwell time scales as $N^3$ for Rouse dynamics and as
$N^{2.5}$ for Zimm dynamics \cite{sungpark2}. These theories indeed
provide insight into the process of translocation, but their usage of
the (equilibrium) free energy to determine the transition rates from
$m$ to $m\pm1$ implicitly assumes that at a fixed reaction coordinate
$m$ the polymers equilibriate much faster than the typical time for
the reaction coordinate to change its value by $\pm1$. This is not
necessarily the case for longer polymers or polymers in higher spatial
dimensions \cite{kantor}. Verifying the scaling results of these
theories using simulations, too, remains a computationally significant
challenge since it involves simulating long polymers and
correspondingly long time scales.

The purpose of this letter is (a) to report the results of extensive
lattice-based Monte Carlo simulations of the unbiased translocation
process in three spatial dimensions, for a variety of polymer lengths,
pore diameters and membrane thickness; and (b) to trace the physical
origin of their differences from the existing mean-field theory
results. Our system consists of two cells A and B, each of volume
$V$, that are connected by a pore of diameter $b$ in a membrane of
thickness $L$. The polymer is modeled as a lattice polymer of $N$
monomers, obeying self-avoiding walk statistics. Its movement
consists of single monomer jumps to neighboring lattice sites. Jumps
along the contour of the polymer, i.e., reptation moves, are attempted
with a higher frequency than jumps that displace the contour of the
polymer laterally to cause Rouse dynamics.  A detailed description of
this lattice polymer model, its computationally efficient
implementation and a study of some of its properties and applications
can be found in Ref.\ \cite{heukelum03,LatMCmodel}. Hydrodynamical
interactions are not incorporated in this model.

In our simulations, the polymer repeatedly moves back and forth from
one cell to the other through the pore [see \fig~\ref{fig2}(a)]. Our
primary interest lies in the scaling behaviour of two quantities, (i)
the {\it mean translocation time\/} $\langle\tau_t\rangle$, the time
required for the whole polymer to escape from one cell to the other,
and (ii) the {\it mean dwell time\/} $\langle\tau_d\rangle$, the time
that the polymer spends in the pore during the translocation
process. To define these quantities precisely, we introduce the
following states of the polymer. In states A and B, the entire polymer
is located in cell A, resp.\ B. States M and ${\text M}'$ are defined
as the states in which the middle monomer is located halfway between
both cells. Finally, states T and ${\text T}'$ are the complementary
to the previous states: the polymer is threaded, but the middle
monomer is not in the middle of the pore. The finer distinction between
states M and T, resp.\ ${\text M}'$ and ${\text T}'$ is that in the
first case, the polymer is on its way from state A to B or vice versa,
while in the second case it originates in state A or B and returns
to the same state. The translocation process in our simulations can
then be characterized by the sequence of these states in time (see
\fig~\ref{fig2}). In this picture, the dwell time is the mean time that
the polymer spends in states M or T, while the translocation time is
the mean time starting at the first instant the polymer reaches state
A after leaving state B, until it reaches state B again.

Having set both Kuhn length of the polymer and the lattice spacing to
unity, we conjecture that for thin membranes ($L\simeq1$)
$\langle\tau_d\rangle\sim N^{1+2\nu}F(b/R_g)$, and verify it for $b=1$
using polymer lengths up to $N=1200$.  For narrow pores ($b\simeq1$),
we argue and verify that $\log \langle \tau_t \rangle \sim L$. We also
observe that $\langle\tau_t\rangle\sim
VN\,(b/R_g)^{-(1+2\nu+\gamma-2\gamma_1)/\nu}F(b/R_g)$.  Here
$\nu=0.588$ is the growth exponent for self-avoiding walks, and
$\gamma=1.1601$ and $\gamma_1=0.68$ are exponents related to the
entropy of a polymer in bulk and near a rigid wall respectively, and
$F(\xi)$ is a scaling function; it approaches a constant for
$\xi\rightarrow\infty$ and behaves $\sim\xi^{-0.38\pm0.08}$ as
$\xi\rightarrow0$.

{\it Argument for the scaling of $\langle\tau_d\rangle$:} For thin
membranes ($L \simeq 1$), a scaling relation between
$\langle\tau_d\rangle$, $b$ and $N$ can be obtained by the following
hypothesis: $\langle\tau_d\rangle=N^\alpha F(b/N^\beta)$ for some
$\alpha$ and $\beta$. We expect that for $\langle\tau_d\rangle$,
$b/R_g$ is a relevant dimensionless parameter that determines how
easily the polymer can squeeze itself through the pore, since the
polymer can ``feel'' the presence of the pore only if its radius of
gyration $R_g$ is comparable to the pore diameter $b$. This implies
that $b/N^\beta\sim b/R_g\Rightarrow\beta=\nu$, as $R_g=\lambda
N^{\nu}$. Moreover, from physical grounds, such a scaling hypothesis
means that the scaling function $F(\xi)$ should approach a constant
for $\xi\to\infty$. Since for very large pores (i.e., $b \gg R_g$) the
polymer no longer feels the pore, $\langle\tau_d\rangle$ should be the
time taken by the polymer to diffuse a distance
$R_g\Rightarrow\langle\tau_d\rangle\sim R_g^2/(2D_N)\sim R_g^2N\sim
N^{1+2\nu}$, where $D_N \sim N^{-1}$ is the diffusion coefficient of
the polymer in a dilute polymer solution. The last relation, together
with the scaling hypothesis implies that $\alpha=1+2\nu$, and
\begin{equation}
\langle\tau_d\rangle \sim N^{2\nu+1}F\left(\frac{b}{R_g}\right)\,.
\label{taudscaling}
\end{equation}

Additionally, since the monomers within the pore move along the
contour of the polymer, i.e., reptate, $\langle\tau_d\rangle$ should
be independent of $L$, as long as $L\ll N$ \cite{kardararg}.

{\it Relation between $\langle\tau_d\rangle$ and $\langle\tau_u\rangle$:}
During the dwelling process, the polymer necessarily has to pass state
M at least once. Due to the spatial symmetry between cells A and B,
each time sequence as depicted in \fig~\ref{fig2} is equally probable
under exchange of states A and B. Additionally, each time sequence from
A to B is as likely as its time-reversed counterpart.

To devise a computationally cheaper method to measure
$\langle\tau_d\rangle$ using these symmetries, we introduce an additional
time-scale $\langle \tau_u \rangle$, the mean unthreading time, which
is the average time that either state A or B is reached from state M
(not excluding possible reoccurrences of state M). Due to time symmetry,
the mean time passed since the polymer last left state A, until it
reached state M, is as large as the mean time passed since the polymer
last left state B, until it reached state M. Consequently,
\begin{equation}
\langle \tau_d \rangle=2 \langle \tau_u \rangle.
\label{eq:taudfromtauu}
\end{equation}
For $b>1$, we expect a similar relation between $\langle \tau_d
\rangle$ and $\langle \tau_u \rangle$ to hold. However, we do not have
a suitable argument for it, since for larger pores, the properties of
the polymer as a macromolecule start to play a role \cite{middle2}.

{\it Relation between $\langle\tau_t\rangle$ and
$\langle\tau_d\rangle$ for $b=1$:} The fraction of time spent in
states M and T compared to $\langle\tau_t\rangle$ equals
$\langle\tau_d\rangle/\langle\tau_t\rangle$.  The probability that the
polymer is threaded exactly halfway is an equilibrium property; hence,
the sum of the probabilities $p_{\text M}$ and $p_{\text M}'$ that the
polymer is in state M or ${\text M}'$ can be obtained from the
contribution of these states to the total partition sum $Z_{\rm
tot}$. The ratios $f_{\text M} \equiv p_{\text M}/p_{{\text M}'}$ and
$f_{\text T} \equiv p_{\text M}/p_{\text T}$ are non-equilibrium
properties, but as we show below, it is possible to estimate these
ratios accurately from targeted simulations. With these quantities,
the average translocation time can be obtained indirectly from the
dwell time, using

\begin{equation}
\langle\tau_t\rangle=\langle\tau_d\rangle \frac{f_{\text T}
(1+f_{\text M})}{(p_{\text M}+p_{\text M'}) f_{\text M} (1+f_{\text
T})}\,.
\label{eq:tautfromtaud}
\end{equation}

We verify Eqs.~({\ref{taudscaling}-\ref{eq:tautfromtaud}) and
cross-check their consistency using direct simulations to measure
$\langle\tau_t\rangle$ and targeted simulations to measure
$\langle\tau_u\rangle$, $p_{\text M}+p_{{\text M}'}$, $f_{\text M}$
and $f_{\text T}$. We use the lattice polymer model of
Ref.~\cite{heukelum03}.

First, we estimate the entropic penalty paid by the polymer in state M
or $\mbox{M}'$. The partition sum $Z_b(N)$ of a self-avoiding polymer
of length $N$, anchored at the origin of an infinite lattice, is
proportional to $Z_b(N) \sim \mu^N N^{\gamma-1}$, with $\gamma=1.1601$
and $\mu$ is a lattice-dependent non-universal constant. If this
polymer is restricted to the half-space $z\ge 0$, the same expression
holds, but with an adjusted exponent $\gamma_1=0.68$ \cite{Diehla98}.
The partition sum of a polymer of length $N$ in state M or
$\mbox{M}'$, threaded through a narrow pore ($b=1$) in a thin membrane
($L=1$) is then given by the product of the partition functions of two
separate self-avoiding polymers of length $N/2$, each having  one of
their ends anchored at a rigid wall, as $Z_t(L=1,N/2,N/2)=\left[
\mu^{N/2} (N/2)^{\gamma_1-1} \right]^2 \sim Z_b(N)
N^{-\gamma+2\gamma_1-1}$.  Adding a linear scaling with $V$ to
$Z_b(N)$, this ratio is also the equilibrium probability that the
polymer is in state M or $\mbox{M}'$. Hence, we obtain
\begin{equation}
p_{\text M}+p_{{\text M}'}(N)\sim N^{-\gamma+2\gamma_1-1}/V.
\label{eq:pM}
\end{equation}

With increasing membrane thickness $L$, since the lattice coordination
number in the pore is much smaller than in the bulk, we have $\log
Z_t(L,N) \sim L$, at least as long as $L \ll N$ (neglecting
logarithmic terms). Stated differently, the entropic barrier
encountered by the translocating polymer increases linearly with
$L$. This results in an exponential increase for $\langle \tau_t
\rangle$ [see Fig. \ref{taut2}]:
\begin{equation}
\log \langle \tau_t \rangle \sim L.
\label{e4}
\end{equation}

Next, we perform dynamical simulations to determine $f_{\text M}$,
$f_{\text T}$ and $\langle \tau_u \rangle$.  The simulations start
with a polymer of length $N$, threaded halfway in the pore, and the
polymer originates from cell A. We then wait until the polymer
unthreads. If it unthreads into cell A, the starting configuration is
labeled $\text{M}'$, while if it unthreads into cell B, it is labeled
M.  We record the ratio of the number of polymers unthreading into
cell A vs. cell B for polymer lengths $N=20, 40, 50$ and 80, and
obtained $1.37, 1.28, 1.33$ and $1.27$ respectively for this ratio;
i.e., cell A is preferred above cell B by a small factor, which does
not increase with $N$. This asymmetry can be easily explained by the
fact that during the translocation process from state A to state B,
the polymer accumulates folded segments on the B side and stretched
segments on the A side, which makes it more prone to go back to state
A than to proceed to state B. Thus, with $c\simeq1.3^{-1}$, we
conclude that
\begin{equation}
f_{\text M} (N) \sim c.
\label{eq:fM}
\end{equation}

In the same simulations, we measured $f_{\text T}/(1+f_{\text T})$,
i.e., the fraction of time that the polymer is in state M before
unthreading. From the theoretical ratio of
$\sum_{i=1}^{N-1}Z_t(L=1,i,N-i)$ and $Z_t(L=1,N/2,N/2)$, we found (and
numerically observed as well) that
\begin{equation}
f_T (N) \sim N^{-1}.
\label{eq:fT}
\end{equation}
\begin{figure}[!h]
\begin{center}
\includegraphics[width=0.6\linewidth,angle=270]{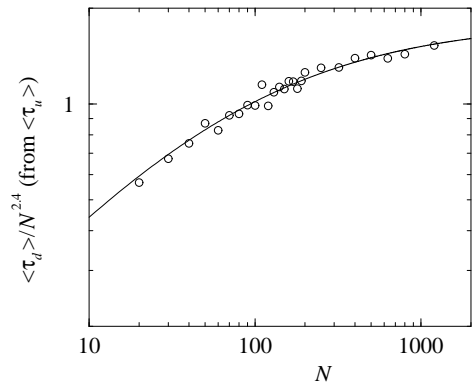}
\end{center}
\caption{$\langle\tau_d\rangle$ obtained from $\langle\tau_u\rangle$
for $L=b=1$ and $N$ up to 1200. Solid line:
$\langle\tau_d\rangle=[0.55N^{-3}+6.84N^{-2.4}]^{-1}$. When the same
data are plotted in the $(b/R_g)$-$(\langle\tau_d\rangle/N^{1+2\nu})$
coordinates, the $\xi^{-0.38\pm0.08}$ scaling for $F(\xi)$ is recovered
for $\xi\rightarrow0$.
\label{fig4}} 
\end{figure}

For each $N$, we combined the unthreading times into a histogram.
We then obtained $\langle \tau_u \rangle (N)$ from a fit of the
long-times tail of this histogram (see \fig~\ref{fig4}). We found that
for short polymers, $\langle \tau_u \rangle (N) \sim N^3$, while for
long polymers, $\langle \tau_u \rangle (N) \sim N^{2.4\pm0.05}$. The
explanation of this is as follows: a polymer translocating from A to B pulls
on its segments in A as it accumulates folded segments in B. The
resulting strains can be released by reptation moves (along the
contour) that initiate {\em only at the ends}, or by Rouse moves
(perpendicular to the contour) that take place {\it anywhere on the
polymer}. For the first mechanism, the scaling of $\langle \tau_u
\rangle$ with $N$ is that of reptation [i.e., $\langle \tau_u \rangle
(N) \sim N^3$, which is the same as the mean-field theory result], but
since there are of $O(N)$ more segments on the polymer (where Rouse
moves can occur) than the two ends (where reptation moves initiate),
the second one dominates for long polymers, giving rise to the
crossover seen in Fig. \ref{fig4}. The precise location of this
crossover depends on the details of the experiment/simulation.

Having combined Eqs.~(\ref{eq:taudfromtauu}-\ref{eq:pM}) and
(\ref{eq:fM}-\ref{eq:fT}), we obtain, for $b=1$,
$\langle\tau_t\rangle$ scaling as a function of polymer length as
\begin{equation}
\langle \tau_t \rangle \sim V
N^{2+2\nu+\gamma-2\gamma_1+0.22\pm0.05}\,.
\label{tautscaling}
\end{equation}
For $b>1$ and $L=1$, a scaling relation for $\langle\tau_t\rangle$ can
be obtained if we assume that $\langle\tau_t\rangle$ is related to
$\langle\tau_d\rangle$ in the same way as in
Eq. (\ref{eq:tautfromtaud}). In this case we expect the entropic
penalty paid by the polymer in state M or M$'$ to behave as a function
of $b/R_g$ --- as explained above Eq. (\ref{taudscaling}), this is the
quantity that determines to what extent the polymer ``feels'' the
presence of the pore, i.e., we expect $N^{-\gamma+2\gamma_1-1}$ in
Eq. (\ref{eq:pM}) to be replaced by
$(N^\nu/b)^{(-\gamma+2\gamma_1-1)/\nu}$. Simultaneously, $V$ in
Eq. (\ref{eq:pM}) is to be replaced by $V/b^2$, as the chance for the
polymer to find the pore increases linearly with the pore area. We
however expect Eqs. (\ref{eq:fM}) and (\ref{eq:fT}) to remain
unchanged as they only concern a threaded polymer. Together with
Eq. (\ref{taudscaling}), for $b>1$ and $L=1$ this argument leads one
to
\begin{eqnarray}
\langle\tau_t\rangle\sim
VN\,(b/R_g)^{-(1+2\nu+\gamma-2\gamma_1)/\nu}F(b/R_g)\,.
\label{tautscalingwithb}
\end{eqnarray}

We performed {\em direct simulations\/} to verify
Eq. (\ref{tautscalingwithb}) [see Fig. \ref{taut2}]. We started with a
polymer in cell A, as shown in \fig~\ref{fig2}, and recorded the times
(for up to 500 different runs) it took to reach state B. We then made
a histogram of these times and deduced $\langle \tau_t \rangle$ from
its asymptotic slope.

In summary, we studied unbiased polymer translocation for various pore
diameters $b$ and membrane thicknesses $L$, using a lattice polymer
model. We found that for $L=1$, both the mean translocation time
$\langle\tau_t\rangle$ and the mean dwell time $\langle\tau_d\rangle$
obey scaling relations that involve functions of $b/R_g$, where $R_g$
is the radius of gyration of the polymer. We also showed that for
$b=1$, $\ln\langle\tau_t\rangle\sim L$ and $\langle\tau_d\rangle$ is
independent of $L$. We explained these results using simple arguments
based on the polymer's dynamical and equilibrium properties. Our
analysis explains, for the first time, how and why deviations from the
mean-field theory results occur for long polymers.
\begin{figure}[h]
\begin{center}
\includegraphics[width=0.7\linewidth]{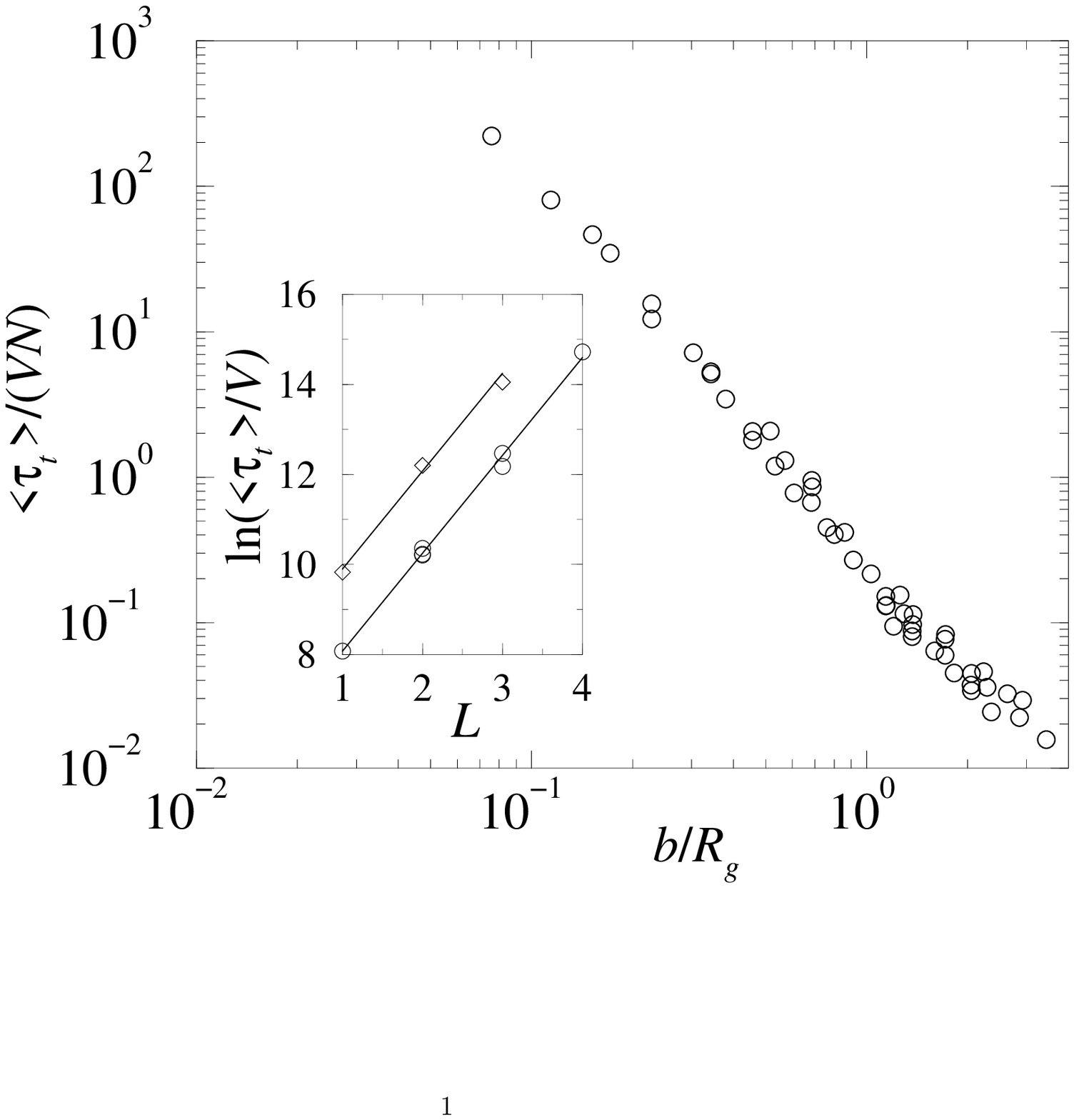}
\end{center}
\caption{$\langle\tau_t\rangle$ vs. $b/R_g$ for $N=20$,
$40$, $80$; $b=1,\ldots,31$. Inset:
$\ln\langle\tau_t\rangle\sim L$ for $b=1$ (circles: $N=20$,
diamonds: $N=40$).} 
\label{taut2}
\end{figure}

The persistence length of a polymer in translocation experiments is
equivalent to the Kuhn length used here. In experiments electric field 
effects due to the applied bias voltage and hydrodynamical effects 
\cite{AC} are always present; these we did not consider here.
Nevertheless, as our scaling results are based on very general
grounds, we expect the {\em same\/} scaling forms (involving $b/R_g$
for $L=1$, or $\ln\langle\tau_t\rangle\sim L$ for $b=1$) to hold when
the hydrodynamical effects are taken into account, albeit with {\it
different\/} exponents. The effects of hydrodynamics and external
fields on translocation are two major topics of our onging work.

We thank Profs.\ Henk van Beijeren and Erich Eisenriegler for helpful
discussions.


\begin{thebibliography}{05}
\bibitem{middle2} For $b>1$, in state M or ${\text M}'$, the two ends of
the polymer may end up in the same cell. How to account for such states
is not clear to us.
\bibitem{AC} In translocation experiments, pore blockage can be detected
with short AC (instead of DC) pulses; this eliminates the effect of the
external fields.
\end{thebibliography}
\end{document}